\documentclass[twocolumn,secnumarabic,amssymb, nobibnotes, prb]{revtex4}
\usepackage{bm}%
\usepackage{graphicx}
\usepackage{subfigure}
\usepackage{multirow}
\usepackage{color}

\begin{document}

\title{\emph{Ab initio} study of the interface properties of Fe/GaAs(110)}%
\author{A. Gr\"unebohm}%
\email{anna@thp.uni-duisburg.de}
\author{H. C. Herper}
\author{P. Entel}
\affiliation{Universit\"at Duisburg-Essen, Lotharstr.\,1, 47048 Duisburg}
\date{March, 2009}%

\begin{abstract}
We have investigated the initial growth of Fe on GaAs(110) by means of
density functional theory. In contrast to the conventionally used
(001)-surface the (110)-surface does not reconstruct. Therefore, a
flat interface and small diffusion can be expected, which makes
Fe/GaAs(110) a possible candidate for spintronic applications. Since experimentally,
the actual quality of the interface seems to depend on the growth conditions, e.g., on
the flux rate, we simulate the effect of different flux rates by
different Fe coverages of the semiconductor surface. Systems with low
coverages are highly diffusive. With increasing amount of Fe, i.e.,
higher flux rates, a flat interface becomes more stable. The magnetic
structure strongly depends on the Fe coverage but no quenching of the
magnetic moments is observed in our calculations. 
 \end{abstract}
\pacs{Valid PACS appear here}
 \maketitle
\section{Introduction}
After the discovery of the giant magnetoresistance GMR \cite{Fert,Gruenberg} and tunnel magnetoresistance TMR effect,\cite{Julliere} spintronics is one of the new fields with high potential for technological applications. The possibility of spin polarization as an additional degree of freedom in charge transport through semiconductor devices gives rise to promising applications, e.g., magnetic random access memory \cite{MRAM} or spin transistors.\cite{spinvalve} 
The Fe/GaAs system is an important candidate for spintronic applications and one of the best studied model systems because of high Curie-temperature of Fe and cheap preparation of the layered systems.
The small lattice mismatch of about 1.24\,\% \cite{chambers} 
 between the GaAs substrate and bcc Fe allows to grow an unstrained interface which is known to be one of the best ordered and most abrupt metal-semiconductor interfaces.\cite{ Kim2000847,godde,Qian}
In addition, the formation of a Schottky barrier at the interface circumvents the conductivity mismatch problem for the hybrid system.\cite{Schmitt,Rashba} 
Until now, many experimental and theoretical investigations have been performed, mostly concerning (001)-oriented systems.\cite {Zhu,mirbt03,Erwin,demchenko:115332, Hong} \\
The (001)-oriented system is experimentally easier to prepare {\color{black} compared to the (110)-oriented surface \cite{li}. However, it is known that GaAs(001) exhibits various surface reconstructions with different terminations depending on the preparation and growth conditions. Such complex surface structures are not expected in case of the non-polar (110)-surface. Therefore we concentrate here on Fe/GaAs(110) although the theoretical investigation is computationally more demanding because twice as many atoms are required in the calculation of the ideal interfaces without reconstructions.}\\
Calculations of ballistic transport properties of Fe/GaAs(001) predict high spin polarization if ideal interfaces are assumed. The large spin injection is related to a matching symmetry of the band structure in (001)-direction, which may act as spin filter.\cite{Mavropoulos} However, the measured spin injection for the (001)-direction varies between 2 and 32\,\%.\cite{Zhu,li} 
Possible reasons for this discrepancy are interdiffusion processes at the Fe-GaAs interface, the formation of intermediate Fe$_x$Ga$_y$As$_z$ layers in the vicinity of the interface,\cite{Rahmoune} or the formation of antiferromagnetic FeAs alloys.\cite{Erwin} The occurrence of these negative
effects seems to depend on the growth conditions. Also, the spin filter effect is destroyed by the ionic relaxations.
Until now, a detailed understanding of the relation between the interface properties, alloy formation and the wide-spread of measured values of spin injection coefficients is lacking. 
At least, it is obvious that the interface structure modifies the electronic structure and thereby the spin polarization at the Fermi level as well as the magnetic moments. \\
For the GaAs(001) interface additional problems arise because of the different surface reconstructions and termination of the GaAs substrate, which influence the interface morphology. 
In contrast to the (001)-surface, no reconstructions are observed for the stoichiometric GaAs(110) surface,\cite{godde,Qian} because of this, it has recently attracted much attention.\cite{igor,winking:193102,Kim2000847,li,godde} But, like for the Fe/GaAs(001) system, the occurrence of interdiffusion seems to depend on surface preparation for the Fe/GaAs(110) system, too.\cite{Ruckman,winking:193102, godde} Ruckman {\it{et al.}} observed an intermixing of Fe with the GaAs(110) substrates for cleaved surfaces which were used without further cleaning.\cite{Ruckman} Recently, Winking {\it {et al.}} reported that Fe/GaAs(110) interfaces can be grown with high crystallinity and rather flat interfaces.\cite{winking:193102} This shows that further investigations are necessary to find optimized growth conditions. \\
The ballistic spin injection in Fe/GaAs(110) is smaller compared to the (001)-oriented system because the matching of the bandstructure
is less perfect. However, a spin injection of at least 13\% has been measured for the Fe/AlGaAs(110) system,\cite{li} meaning that spin polarization and injection coefficients may be
slightly smaller but may more reliably be reproduced. Hence, the (110)-oriented system is a promising candidate for applications requiring cheap materials and easy preparation.\\
Until now, a systematic theoretical investigation of the Fe/GaAs(110) interface structure and its growth process is lacking. Also, no theoretical study of interdiffusion processes at the interface and related magnetic and electronic properties has been performed so far.\\
In this paper we investigate the interface structure of the Fe/GaAs(110) system by simulating different growth conditions. To be more specific, we calculate the energy surface for different Fe configurations for the case of one quarter of an Fe monolayer on the fully relaxed free GaAs(110) surface. This serves as a tool to simulate low growth rates during the fabrication process of the first monolayers. In a second approach, we study the effect of a larger Fe flux during the growth process by increasing the number of Fe atoms. Therefore, we investigate different numbers of Fe layers on the substrate, as well as periodically repeated GaAs/Fe/GaAs multilayers. The interface roughness, relaxation effects, {\color{black}electronic structure} and magnetism for these setups are discussed with respect to the interface structure.
\section{Computational details}
The electronic structure has been calculated self-consistently by using first-principles density functional theory and the plane wave pseudopotential code VASP.\cite{Kresse1} The Projector augmented wave potentials \cite{Blochl} have been used, employing the generalized gradient approximation in the formulation of Perdew, Burke and Ernzerhof \cite{PBE} for the exchange correlation potential. An energy cutoff of 334.9\,eV and a set of $17\times11\times5$ k-points constructed with the Monkhorst-Pack scheme have been used.\cite{Monkhorst} 
{\color{black}The linear tetrahedron sampling has been used for all calculations aside from the  density of states, for which a Gauss-smearing of 0.1\,eV has been used.}
 All energies were converged with an accuracy of $10^{-7}$\,eV. Local magnetic moments were obtained by projecting the wave functions onto spherical harmonics within spheres with radii $r_{Ga, As}=1.217$\,{\AA} and $r_{Fe}=1.302$\,{\AA}.\\
In order to compare energies of slabs containing different numbers of atoms, the formation energy
\begin{equation}
E_{form}=E_{tot}-\sum_iN_i\mu_i
\label{eq:fomr}
\end{equation}
is calculated from the number of nonequivalent atoms $N_i$ of element $i$ and corresponding chemical potentials $\mu_i$. The chemical potentials have been estimated from bulk calculations of bcc Fe and zincblende GaAs, respectively.
Since the small overestimation of the lattice constant by GGA leads to an increased underestimation of the bandgap, we used the experimental lattice constant of GaAs, $a=5.654$\,{\AA} in all calculations. In order to investigate relaxation effects of free surfaces, one side of the slab was passivated with pseudo-hydrogen \cite{pseudohyd} and the lowermost Ga and As ions were fixed at their bulk positions while the  other ions were allowed to relax.
At least 10\,{\AA} of vacuum were used to prevent interactions between the periodically repeated slabs in case of free surfaces.
The relaxation of the ions was carried out until the forces were converged to 0.01\,{\AA}/eV unlike otherwise stated. Additional calculations including the Hubbard-like U-term for the Ga-\emph{d}-orbitals yield similar results for the energy landscape and are showing only tiny effect on the lattice relaxations. Therefore, U-corrections have been neglected in this work.\\
\section{The GaAs surface}
\begin{figure}
\includegraphics[width=0.35\textwidth]{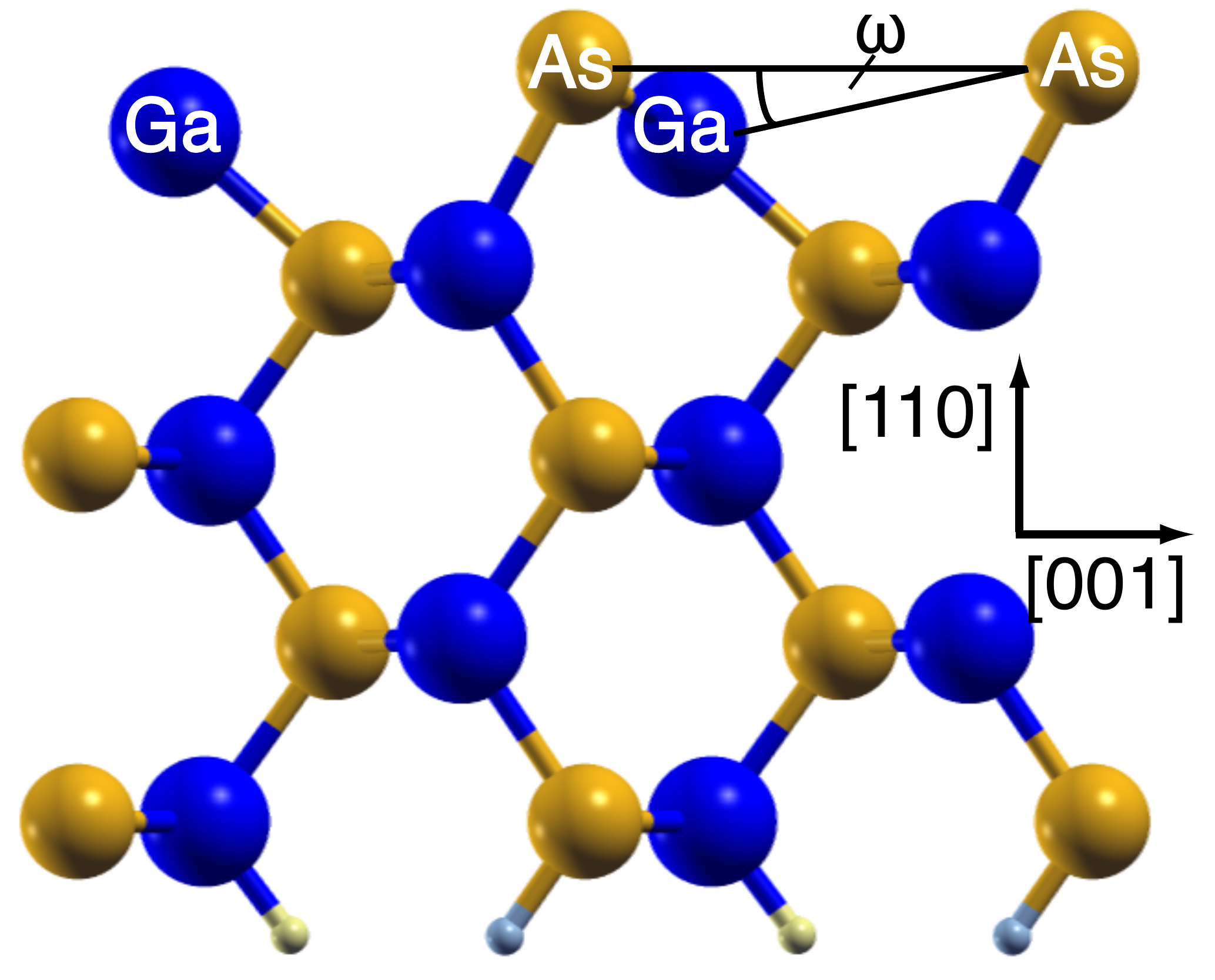}
\caption{The relaxed free GaAs(110) surface. The bottom of the cell is passivated with pseudo-hydrogen, $\omega$ represents the relaxation angle between the surface atoms.} 
\label{fig:superz}
\end{figure}
The GaAs(110) surface does not reconstruct. Instead, the surface energy is lowered through hybridization of the dangling bonds causing relaxation of the surface ions,\cite{Qian} compare Fig.\,\ref{fig:superz}. 
Here, the Ga atoms relax into the substrate, resulting in a $sp2$-hybridization while the As atoms move in opposite direction. During this process, the surface appears buckled. Hereby, no relaxation in \mbox{(-110)}-direction is observed while the Ga and As ions reduce the (001)-component of their distance in order to conserve their interatomic distance despite the buckling, see topview of the relaxed surface in Fig.\,\ref{fig:blubb}. A detailed discussion can be found in Ref.\,\onlinecite{Haugk}.
\begin{figure}
\includegraphics[width=0.35\textwidth]{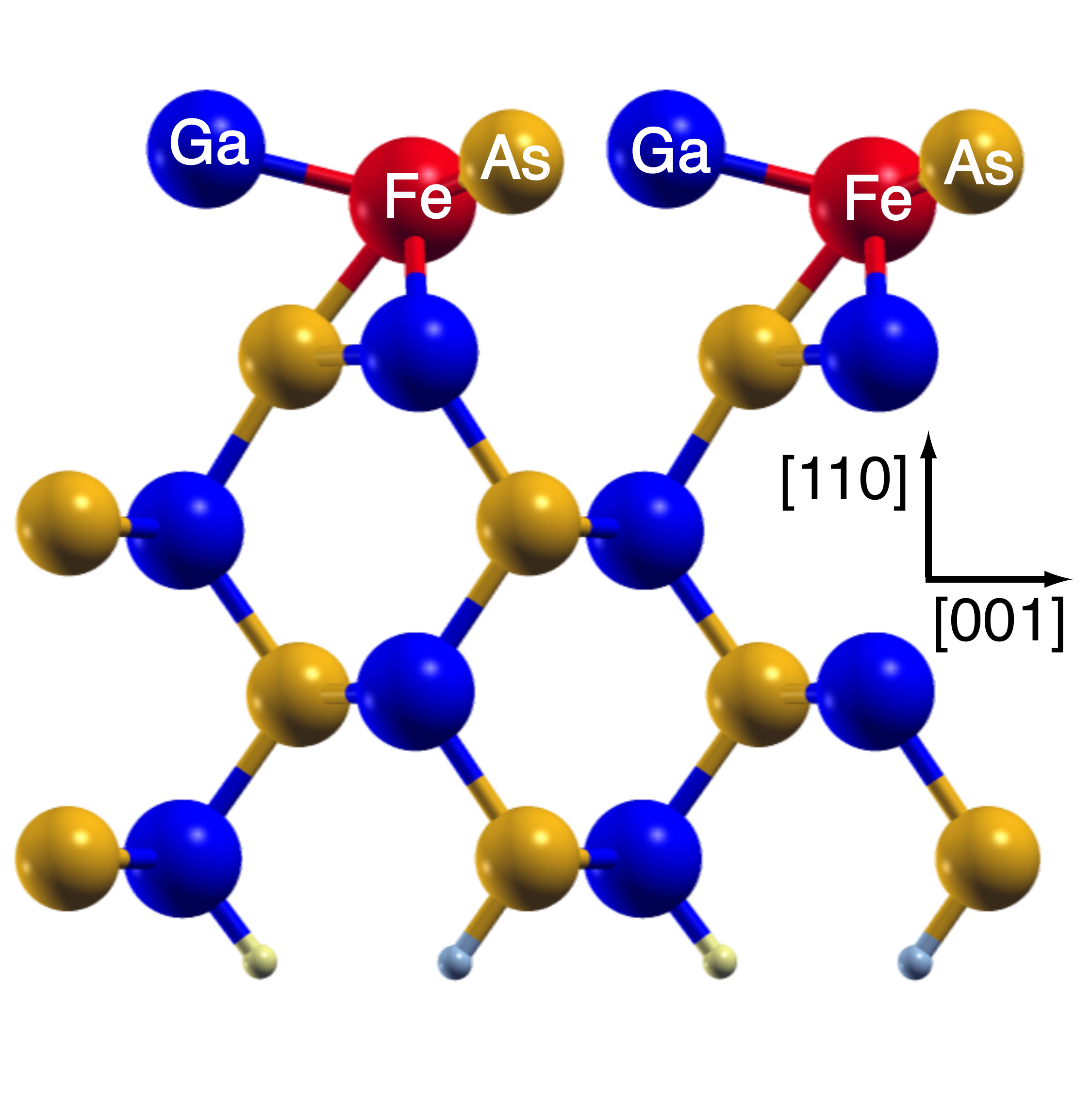}
\caption{Relaxed structure in case of 1/4 monolayer of Fe on the GaAs(110) surface, which is the energetically most favorable configuration 1, see also Fig.\,\ref{fig:blubb}.}
\label{fig:1fe}
\end{figure}
Our calculations yield a buckling angle for the free surface of
\begin{equation}
\omega=\tan^{-1}=\frac{\Delta_{110}}{\Delta_{001}}=31^\circ \,.
\end{equation}
This is in in agreement with earlier experimental and theoretical results.\cite{Qian}
\section{{Fe} adatoms}
\begin{figure}
\begin{center}
\includegraphics[width=0.45\textwidth]{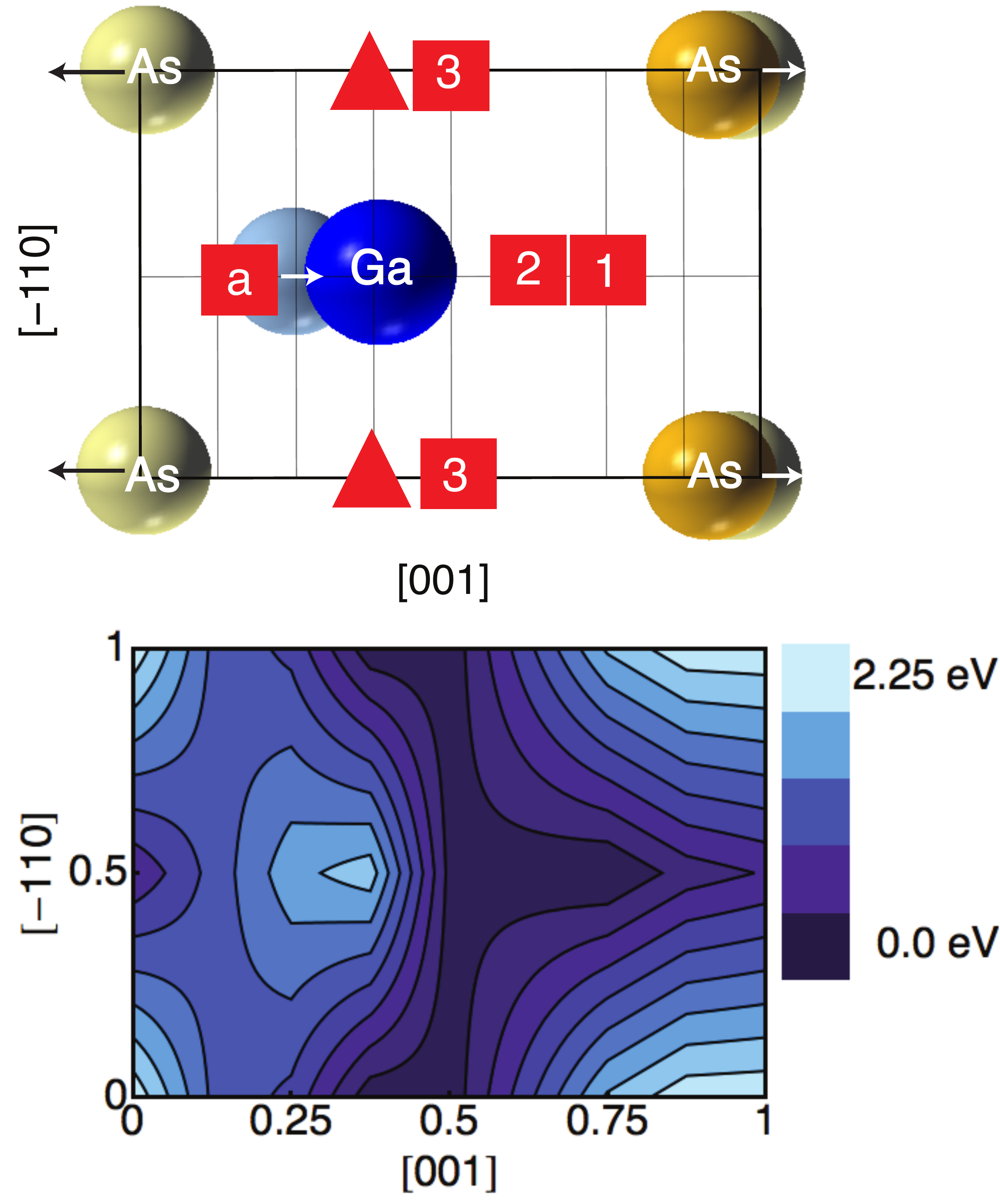}
\caption{Top: Topview of the relaxed GaAs(110) surface. The brighter ions mark the bulk positions of the corresponding atoms while the darker ions illustrate the relaxed position at the free surface. The different investigated Fe positions are indicated by the grid. Numbers label the 3 Fe positions which are most favorable. The triangles mark a metastable state, see text {\color{black} and "a" corresponds to the energetically most favorable configuration with interdiffusion}. Bottom: The energy surface in eV for the  Fe positions on the above shown grid. Contour lines are at a distance of 0.25\,eV.}
\label{fig:blubb}
\end{center}
\end{figure}
\begin{table}
\caption{Nearest neighbor distances for the GaAs(110) surface with 1/4 adlayer Fe for the energetically most favorable interface configurations, see Figs.\,\ref{fig:1fe} and \ref{fig:blubb}. The planar Fe coordinates are given in fractional coordinates of the supercell, see Fig.\,\ref{fig:blubb}. The values in brackets correspond to the number of neighbors. {\color{black} The energies are given relative to the energetic ground state of a system with Fe adatom in meV/atom.}}
\begin{tabular}{ccccccc}
\hline
\hline
{\color{black}Position}&\multicolumn{2}{c}{In-plane Fe }&\multicolumn{2}{c}{Interatomic distance}&{\color{black} Energy}\\
{\color{black}in Fig.\,\ref{fig:blubb}}&\multicolumn{2}{c}{coordinates}&\multicolumn{2}{c}{({\AA})}&{\color{black}(meV/atom)}\\
\hline
&[-110]&[001]&As-Fe&Ga-Fe\\
\hline
\multirow{2}{*}{{\color{black}(1)}}&\multirow{2}{*}{0.5}&\multirow{2}{*}{0.75}&2.386 (2)&2.739 (2)&\multirow{2}{*}{\color{black}0}\\
&&&2.438 (1)&2.615 (1)\\
\hline
\multirow{2}{*}{{\color{black}(2)}}&\multirow{2}{*}{0.5}&\multirow{2}{*}{0.625}&2.443 (2)&-&\multirow{2}{*}{\color{black}1.2}\\
&&&2.412 (1)&2.493 (1)\\
\hline
\multirow{2}{*}{{\color{black}(3)}}&\multirow{2}{*}{0}&\multirow{2}{*}{0.5}&2.590 (2)&2.583 (2)&\multirow{2}{*}{\color{black} 7.4}\\
&&&-&2.441 (1)\\
\hline
\multirow{2}{*}{{\color{black}(a)}}&\multirow{2}{*}{0.5}&\multirow{2}{*}{0.19\footnote{Intermixed surface: Fe at Ga position with Ga adatom}}&2.469 (2)&2.597 (2)&\multirow{2}{*}{\color{black} -31.8}\\
&&&2.421 (1)&-\\
\hline
&\multicolumn{2}{c}{Zincblende (Ref.\,\onlinecite{mirbt03}) }&2.3~~~~(4)&2.4~~~~(4)&{\color{black}-}\\
\hline
\hline
\end{tabular}
\label{tab:abstand}
\end{table}
In order to investigate the initial state of growth of Fe on GaAs(110) under a moderate Fe flux during the growth process, like in Ref.\,\onlinecite{godde}, single Fe adatoms, corresponding to quarter of a monolayer, have been deposited on the relaxed semiconductor surface, see Figs.\,\ref{fig:1fe} and
 \ref{fig:blubb}. The atomic positions within the 4 topmost GaAs layers and the position of the Fe atoms perpendicular to the surface have been relaxed. The in-plane positions of Fe atoms have been kept fixed in the calculation of the energy surface. This constrain hinders the penetration of Fe atoms into the surface, which occurs when Fe is placed directly on top of the As or Ga atoms. These configurations have been neglected in the energy surface shown in Fig.\,\ref{fig:blubb}, because without relaxation of the in-plane coordinates, a sufficient minimization of the forces was not possible.\\ 
From the present calculations it turns out that for such low coverages, the energy of the Fe/GaAs(110) system is lowered if the Fe atoms penetrate into the semiconductor surface, see Fig.\,\ref{fig:1fe}. 
The three energetically most favorable Fe positions are marked by numbers in Fig.\,\ref{fig:blubb}. 
{\color{black} The energy difference between the ground state (labeled by 1) and configuration 2 and 3 is less than 1.2\,meV/atom and 7.4\,meV/atom, respectively. This corresponds to a thermal energy of 14\,K (86\,K), which means that these configurations may be stable at finite temperatures, too.
Here, the small energy contribution of the pseudo-hydrogen atoms to the total energy has been neglected.}\\
The driving force for the Fe relaxation is the low coordination of the adatoms, because larger Fe coordination delivers a large amount of energy. The Fe atoms in these configurations possess at least three As or Ga atoms in their direct neighborhood, see Table\,\ref{tab:abstand}. Due to a larger Fe-$d$-As-$p$ hybridization, the As-Fe interaction is much stronger than the Ga-Fe interaction. Hence, in the ground state the Fe atom has two As neighbors at a distance of 2.386\,{\AA}. Our results for the interatomic distances are in qualitative agreement with calculations on artificial As-Fe (Ga-Fe) zincblende structures, from which an optimal distance of about 2.3\,{\AA} (2.4\,{\AA}) for the As-Fe (Ga-Fe) structure was obtained by Mirbt et al.\cite{mirbt03}\\
In case of the configuration marked by a triangle in Fig.\,\ref{fig:blubb}, the Fe atoms are trapped in a local energy minimum at a distance of 2.446\,{\AA} towards the As atoms and thereby a relaxation with the demanded accuracy for the forces was not possible. Therefore, this value, which lies about 2 eV above the ground-state, was not included in the calculation of the energy surface (Fig.\,\ref{fig:blubb}).\\ 
As a consequence of the energy gain through Fe-As hybridization, the energy of the system is lowered, if the topmost Ga-As bond is broken and the Fe atom replaces a Ga atom. This leads to a Ga adatom which is bonded to the atoms in the surface. The energy gain is in agreement with previous findings for the (001)-surface.\cite{mirbt03,Erwin}
In case of Ga adatoms, the energy is reduced by 0.35\,eV compared to the groundstate with Fe adatoms. The most favorable Fe position in the case of such Ga adatoms is listed in Table\,\ref{tab:abstand}. 
However, no such configuration has been found for As adatoms. In this case the energy is always higher than the ideal interface because the energy gain through Fe-Ga-hybridization is not sufficient to break As-Ga bonds.\\
We have only investigated interdiffusion in the first GaAs layers, following the prediction for (001)-oriented systems by Erwin {\it {et al.}} \cite{Erwin} that Fe atoms are trapped at the Ga positions. \\
In summary, our calculations have shown that strong relaxations of the ions appear during the initial state of growth in case of a small amount of Fe. Depending on the kinetic conditions, interdiffusion of Fe and Ga atoms increase.
This may explain experimental results in Ref.\,\onlinecite{godde}, where a small amount of Fe on the GaAs surface is realized by a small Fe flux. 
Under this growth condition no flat Fe films form. Instead Fe island growth is observed at room temperature which leads to highly intermixed Fe/GaAs surface structures through annealing.
 \section{The ideal interface}
\begin{figure}
\includegraphics[width=0.35\textwidth]{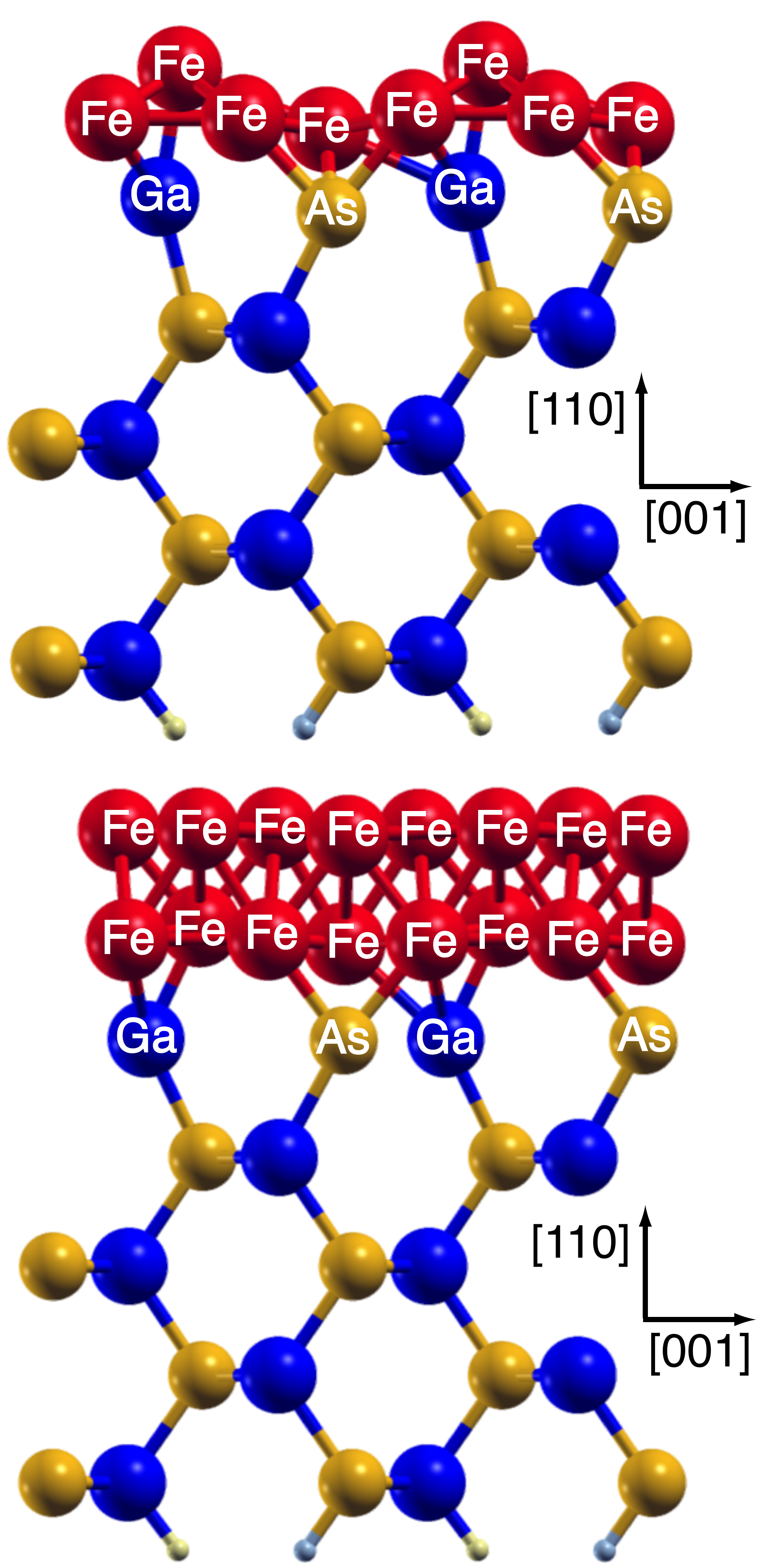}
\caption{Top: Relaxed supercell with a monolayer of Fe on the free GaAs(110) surface. Bottom: Relaxed supercell after a second monolayers of Fe has been placed on the supercell above. The bottom of the cell is passivated with pseudo-hydrogen.}
\label{fig:adlayer}
\end{figure}
\begin{figure}
\begin{center}
\includegraphics[width=0.45\textwidth]{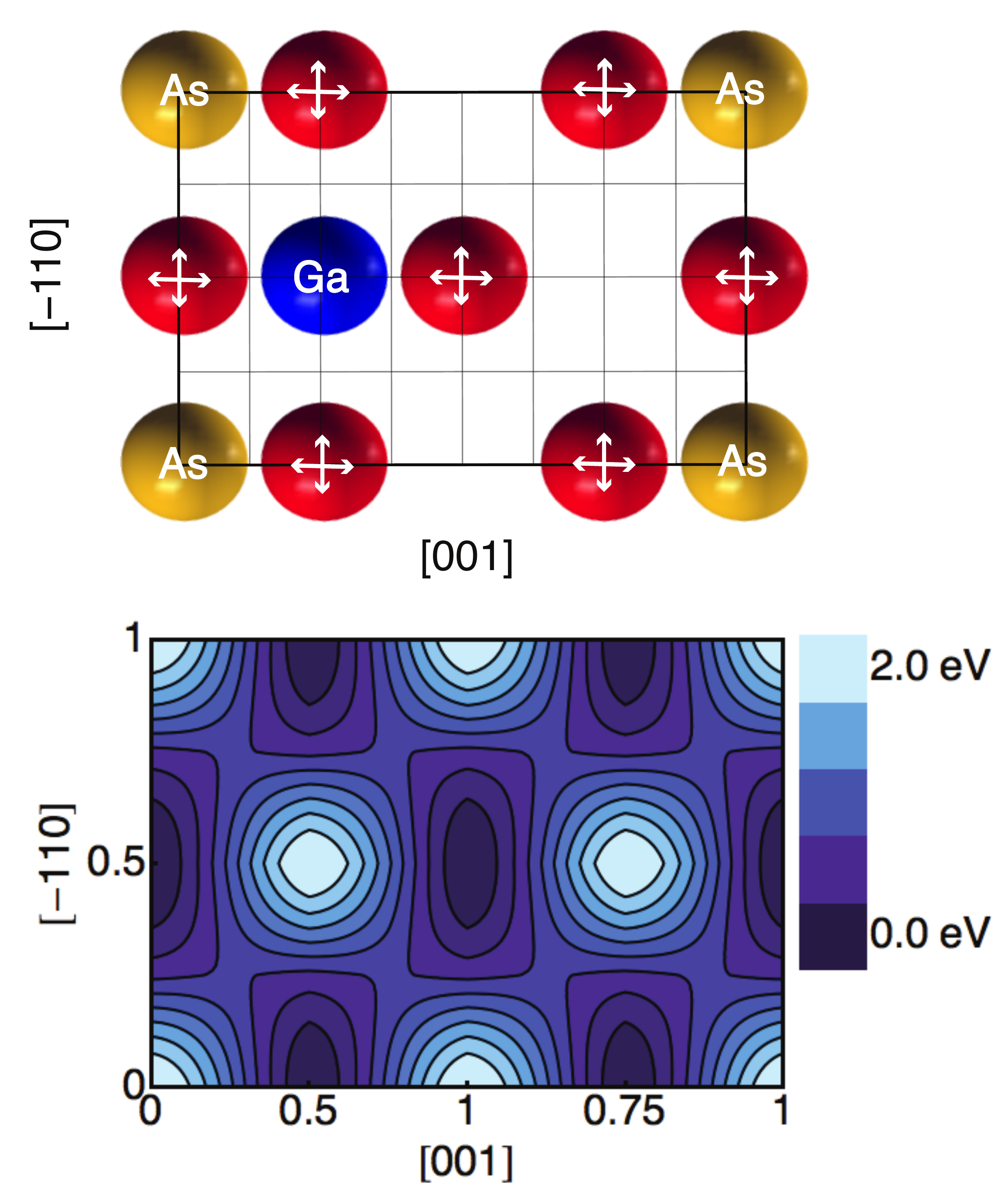}
\caption{Top: Topview of the ideal Fe/GaAs interface
{\color{black} in the most favorable configuration. Additionally, the Fe layer has been shifted rigidly relative to the GaAs surface as indicated by the white arrows. For each grid point the energy has been calculated. Bottom: Energy surface of the ideal Fe/GaAs system for the different configurations defined above.} The energy is given with respect to the ground state configuration in eV (for the supercell with 5 Ga and 5 As atoms as well as 12 Fe atoms). Contour lines are at a distance of {\color{black}0.25\,eV}. 
 {\color{black}The energy surface is calculated for the ideal atomic positions in plane, whereas the ionic positions perpendicular to the interface have been relaxed.}
}
\label{fig:ideal}
\end{center}
\end{figure}
The interface structure discussed in the last section is valid in case of a small Fe flux during the growth process. In order to model a larger flux, a different approach is necessary, because the experimental flux varies between 0.4\,monolayer/min \cite{godde} and 2\,monolayer/min.\cite{winking:193102}
Hence, Fe monolayers have been successively put onto the relaxed GaAs surface. For a qualitative discussion, atomic relaxations were performed until the forces were converged to 0.05\,eV/{\AA}. \\
If one monolayer of Fe is placed onto the GaAs substrate, the relaxation of the free GaAs(110) surface vanishes. The As-Fe and Ga-Fe distances are decreased by the Fe-GaAs interaction.
As a result, the Ga atoms relax outwards and the Ga-As distance is enlarged near the interface, see Fig.\,\ref{fig:adlayer}. 
In contrast to the case of 1/4 monolayer, no relaxation of Fe atoms into the GaAs surface occurs, because the driving force for such relaxations is reduced. The reason for this reduction is the increase of the coordination of the Fe atoms compared to the case of 1/4 monolayer because this coordination is related to an enhanced Fe-Fe interaction which reduces the Fe-GaAs hybridization.
The further growth process is simulated by placing a second monolayer of Fe onto the relaxed Fe/GaAs system.
After atomic relaxation, the Fe-GaAs interface is nearly flat and the 2\,ML of Fe form a bcc like configuration on top of the semiconductor surface, see Fig.\,\ref{fig:adlayer}. This {\color{black} tendency of Fe atoms interpenetrating the surface until a flat surface is formed,} can be attributed to 
 {\color{black} changes of the coordination. On one hand, an increasing coordination of the Fe atoms for the first deposited atoms lowers the energy of the system, i.e., the penetration into the GaAs layer is favorable. On the other hand, with an increasing amount of Fe in the surface layer, the Ga and As atoms would be over-coordinated. Due to this over-coordination, Fe electrons appear in the surface layers which occupy anti-bonding electronic states as all bonding states are already occupied. Therefore,} {\color{black} the bonds become energetically less favorable with an increasing number of Fe atoms inside the GaAs surface. 
Because of this, the relaxation of the Fe atoms into the semiconductor surface is suppressed after the growth of one Fe layer as the enlarged Fe coordination lowers the driving force for the relaxation which is no longer sufficient to overcome the occupation of anti-bonding electronic states. Similar results have been obtained for the (001) growth direction in Ref.\,\onlinecite{Erwin}.}
 \\
It is likely that the presence of an Fe film reduces the Fe mobility and may hinder any further diffusion processes. This plausibility argument, already pointed out in literature,\cite{mirbt03} allows to neglect interdiffusion after the growth of a flat Fe film. The formation of such flat interfaces after two monolayers of Fe in the case of a large Fe flux is in agreement with recent experimental results.\cite{winking:193102} In Ref.\,\onlinecite{winking:193102}, interdiffusion is further suppressed through a low deposition temperature of $T=130$\,K during the growth of the first Fe layers.
 After the formation of a flat interface interdiffusion is no longer energetically advantageous and the flat, abrupt interface is stable under annealing conditions up to 345\,K.\\
In the following, we discuss the properties of thicker Fe layers in the framework of an ideal periodically repeated Fe(110)/GaAs(110) supercell without vacuum, see Fig.\,\ref{fig:ideal}. In this case surface effects are suppressed by the periodic boundary conditions. 
To sample the energy surface, we shifted the GaAs layers {\color{black} rigidly} against the Fe layers on the grid shown in Fig.\,\ref{fig:ideal}.
{\color{black} After the shift the corresponding in-plane coordinates have been fixed whereas the coordinates perpendicular to the interface have been relaxed. This procedure leads to a sampling of the whole energy surface for the different relative GaAs-Fe layer positions, whereas further ionic relaxation in plane would only sample local energy minima.}
 In the ground state, the Fe atoms sit on top of the interstitials of the underlying GaAs layer while Fe atoms on top of the Ga or As positions are most unfavorable.
{\color{black}  For some configurations, e.g., the most favorable configuration, all ions have been relaxed without constrains  to test the quality of the assumed rigid shift between the layers. As this further relaxation has only minor effects on the interface structure and the relative energies of the tested configurations, the rigid shift between the layers can be accounted a proper model for the energy surface.\\
 Nevertheless, all further discussion refer to configurations which were relaxed without constrains as magnetism and density of states are more sensitive to small modifications of the structure than the energy surface.\\} 
Because Fe has twice the number of atoms per layer compared to GaAs, the GaAs cell offers unoccupied positions for the Fe atoms, see Fig.\,\ref{fig:ideal}.
The filling of these positions in the surface layer was found to be stable for the (001)-direction.\cite{demchenko:115332} In analogy, we investigated different configurations with one or two Fe atoms per unit cell in the GaAs(110) interface layer. However, it turned out that the formation of these slabs was at least 1.2\,eV higher in energy compared to the ideal interface, which allows us to conclude that alike Fe diffusion is not stable in case of an (110)-interface, at least at $T=0$\,K. This is mainly due to the stoichiometry of the interface. The strong As-Fe-hybridization at the interface balances the under-coordination of the Fe atoms and the filling of Fe atoms into the pure As interface becomes unstable as already discussed in Ref.\,\onlinecite{demchenko:115332}. The same physics seems to hold for the stoichiometric (110)-interface, which contains one As and one Ga atom per unit cell. 
\section{Magnetic properties}
\begin{table}
\caption{Magnetization of the Fe atoms in $\mu_B$/Fe atom. For the ideal interface an average value is given. Numbers of the Fe position are related to Fig.\,\ref{fig:blubb}.}
\begin{tabular}{ccccc}
\hline
\hline
configuration&Fe position&Magnetic moment [$\mu_B$]\\
\hline
\multirow{2}{*}{ideal interface}&interspace&2.43\\
&on As &2.31\\
\hline
\multirow{4}{*}{1/4 ML}&1&2.62&\\
&2 &2.79\\
&3&2.86\\
&Ga-Fe interdiffusion&2.66\\
&As-Fe interdiffusion&2.20\\
\hline
\hline
\end{tabular}
\label{tab:mag}
\end{table}
The magnitude of the magnetic moments at the interface is one important ingredient for the usability of the material system in spintronic devices.
Former studies have found a large decrease or even a total quenching of the magnetic moments at the Fe/GaAs(001) interface, e.g., Ref.\,\onlinecite{mirbt03}. Therefore, we present here the magnetic properties of the Fe/GaAs(110) interface in detail 
{\color{black}{focusing on two different aspects: first, we investigate the properties of the ideal interface in the framework of a periodically repeated GaAs/Fe/GaAs supercell. Second, we study the influence of ionic relaxations and a free surface on the magnetism of  single Fe atoms on the GaAs surface.}}
\subsection{Magnetism for Fe adatoms}
For 1/4 monolayer of Fe adatoms on the free GaAs(110) surface, no quenching of the magnetic moments has been observed, cf. Table\,\ref{tab:mag}. Although, the As-Fe distance decreases through atomic relaxations no critical value for quenching of the magnetic moments as reported in Ref.\,\onlinecite{mirbt03} has been observed.
Even in the case of an Fe-As distance of 2.3\,{\AA}, which appears for As interdiffusion in the top layer, a finite magnetization appears, cf. Table\,\ref{tab:mag}.
Accordingly, it seems that the (110)-surface is less sensitive to quenching of the magnetic moments, than the (001)-surface. This is partly due to the fact that the GaAs(110) surface is stoichiometric. Pure As interfaces in case of an (001)-orientation may lead to a large As-Fe hybridization and thereby cause a delocalization of the Fe-\emph{d}-states. Here, the Ga atoms in the (110)-configuration reduce this hybridization. To proof this statement, further calculations including non-stoichiometric configurations of the GaAs substrate have to be performed in future.
\begin{figure}
\includegraphics[width=0.5\textwidth]{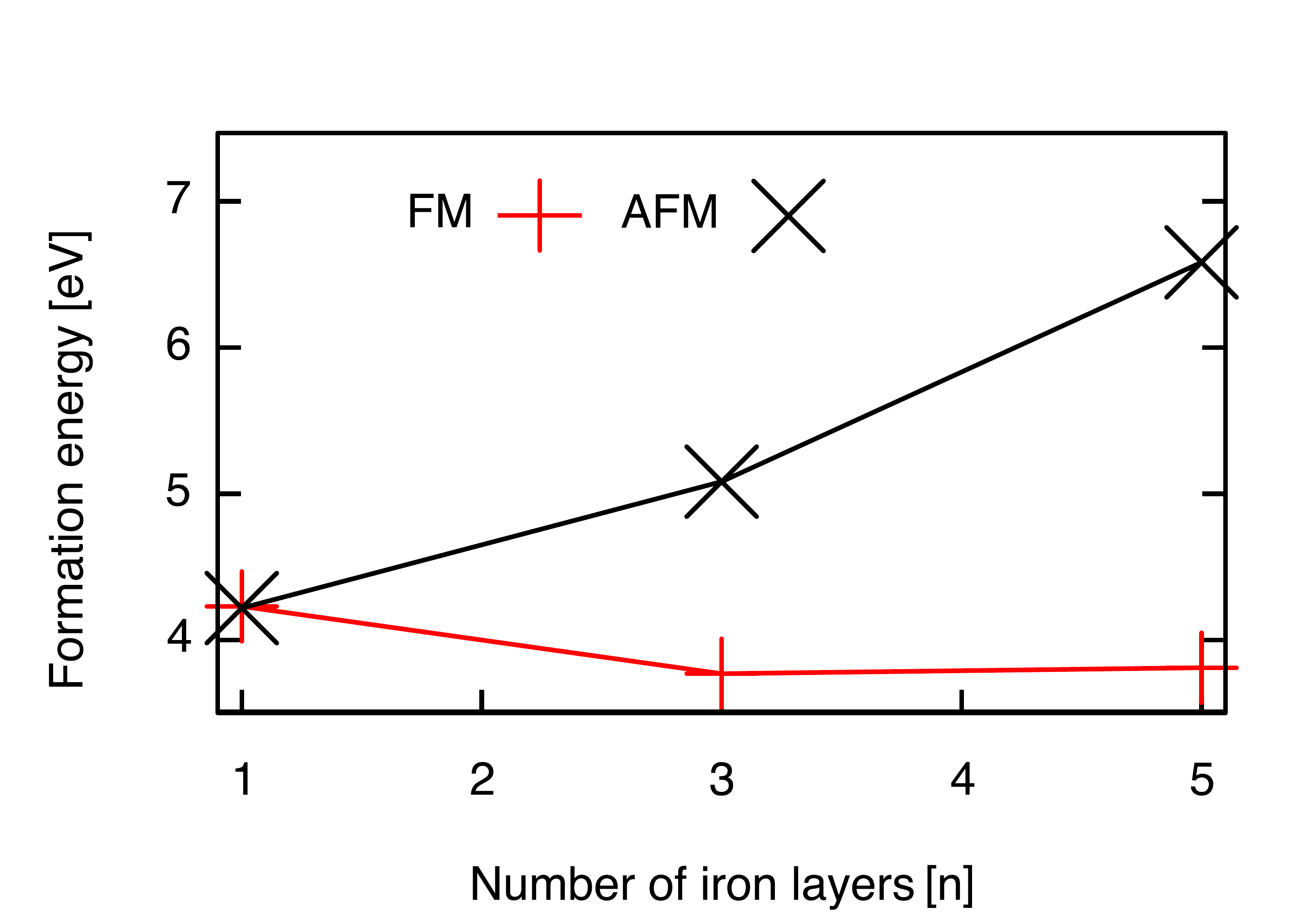}
\caption{{\color{black} Formation energy for GaAs/Fe$_n$/GaAs multilayers. Shown are ferromagnetic and the most favorable antiferromagnetic phase, respectively, compare Fig.\,\ref{fig:afm}. }}
\label{fig:mag}
\end{figure}
\begin{figure}
\includegraphics[width=0.35\textwidth]{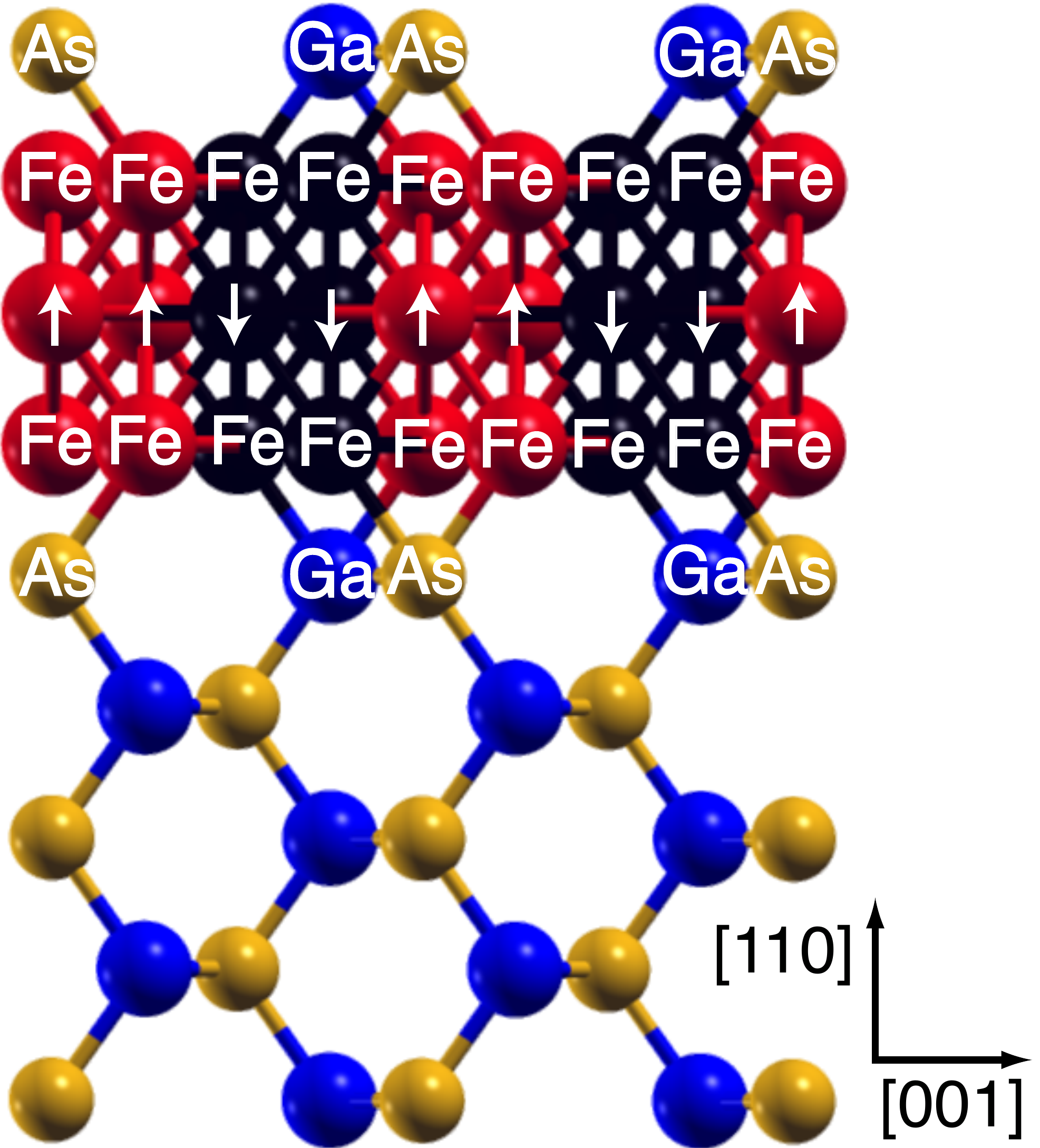}
\caption{{\color{black} Energetically most favorable antiferromagnetic configuration for the periodically repeated Fe/GaAs system. The different colors of the Fe atoms indicate the orientation of the  magnetic moments.}}
\label{fig:afm}
\end{figure}
\subsection{Magnetism for the ideal interface}
{\color{black}Since the structural investigations have shown that with increasing amount of Fe the surface becomes flat, we have studied the magnetic properties of GaAs/Fe/GaAs(110)} multilayers. {\color{black} Here,} no quenching or significant decrease of the magnetic moments of Fe has been observed. For the ground state, the magnetic moments amount to 2.4\,$\mu_B$. The resulting enhancement of 9\,\% compared to the bulk magnetization of Fe {\color{black} mainly} is a surface effect because of the large percentage of Fe at the interface (the Fe moments are enlarged if the number of Fe neighbors is decreased). The same holds for the less stable configurations. Only if the Fe atoms are placed on top of the As atoms, the average magnetic moment is reduced to 2.31\,$\mu_B$ and the magnetic moments of the Fe atoms on top of the As atoms are reduced to 1.9\,$\mu_B$.\\ 
Instead of quenched magnetic moments, the appearance of magnetic inactive layers was attributed to the formation of an antiferromagnetic Fe phase in Refs.\,\onlinecite{Erwin} and \onlinecite{Hong} for one monolayer of Fe on As-terminated GaAs(001). We obtain similar results for one monolayer of Fe in GaAs(110)/Fe/GaAs(110). For this system, the antiferromagnetic Fe phase has the same formation energy as for the ferromagnetic solution, see Fig.\,\ref{fig:mag}. However, with increasing thickness of the Fe layer, the antiferromagnetic phase becomes unstable compared to the ferromagnetic solution. {\color{black}{The corresponding antiferromagnetic phase consists of a domain like antiferromagnetic }} {\color{black}{ordering in each Fe layer, see Fig.\,\ref{fig:afm}. Additional antiferromagnetic phases have also been investigated, but these configurations were found to have a higher formation energy, e.g., a layer by layer change of the direction of the magnetic moments corresponds to a formation energy of 6.9\,eV for three layers of Fe.}}
In contrast to Ref.\,\onlinecite{mirbt03}, we obtain no quenching of the magnetic moments in the antiferromagnetic case. \\
Furthermore, we investigated a possible nonmagnetic phase at the interface, which turns out to be 6.4\,eV higher in energy than the ferromagnetic solution.\\ 
In summary, our results show that high magnetic moments of the Fe atoms at the GaAs(110) interface can be obtained as rather flat interfaces form under adequate growth conditions.
\section{Electronic structure}
{\color{black}{
\begin{figure}
\includegraphics[width=0.45\textwidth]{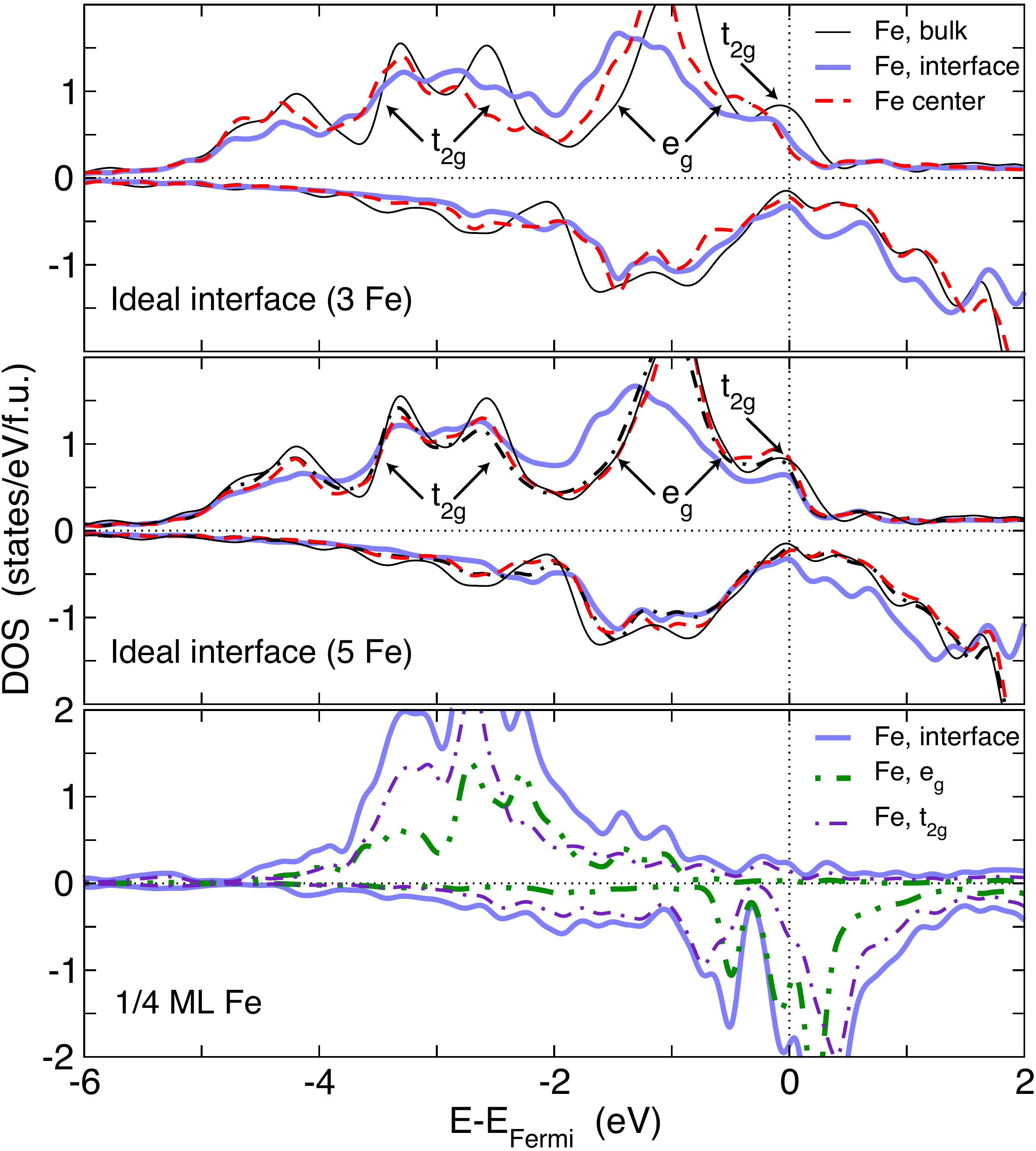}
\caption{Site-projected density of states. Top: Ideal interface for GaAs/Fe$_3$/GaAs. Center: Ideal interface for GaAs/Fe$_5$/GaAs. Bottom: 1/4 monolayer of Fe on the free surface in configuration "a", see Fig.\,\ref{fig:1fe}.}
\label{fig:dos}
\end{figure}
Besides the magnetic moments at the Fe/GaAs(110) interface, the density of state (DOS) close to the Fermi level is another important ingredient for the usability of the material in spintronic devices.
Especially the spin polarization $P$ at the Fermi level
\begin{equation}
P=\frac{N_{\uparrow}(E_F)-N_{\downarrow}(E_F)}{N_{\uparrow}(E_F)+N_{\downarrow}(E_F)}\,,
\end{equation}
where $N_{\uparrow,\downarrow}  (E_F)$  quantifies the DOS at the Fermi level, has a large influence on the spin transport properties. As DOS  and $P$ are sensitive to the interface structure in case of layered systems like Fe/GaAs(110) \cite{icm,intermag}, we will discuss the influence of the different interface configurations in the following.
Figure\,\ref{fig:dos} shows the site-projected density of states for the Fe atoms of different Fe/GaAs(110) systems.
In the upper panels, the DOS for the ideal interface is compared with the DOS of the Fe bulk system, where, the bulk system has been calculated using the same lattice constant. As obvious from Fig.\,\ref{fig:dos}, the main features of the bulk DOS are conserved even at the direct interface. However, some important modifications appear. First, the majority DOS at the Fermi level is slightly reduced. While in bulk Fe, a  \emph{d}-peak with t$_{2g}$-symmetry appears at the Fermi level, this peak is reduced at the GaAs(110) interface.
At the same time the minority DOS at the Fermi level is slightly enhanced. Due to these modifications, the polarization of the DOS at the Fermi level is noticeable reduced, but a finite polarization at the direct interface is conserved for the ideal GaAs/Fe(110)-system, see Ref.\,\onlinecite{intermag} for a detailed discussion.
Also, the reduction of the polarization only appears in the vicinity of the interface. For the GaAs/Fe$_3$/GaAs system, a polarization of 19.6\,\% appears at the direct interface layer whereas the polarization approaches 39.2\,\% in the second Fe layer \cite{intermag} since the minority DOS is as small as in the bulk case, see top panel of Fig.\,\ref{fig:dos}.
For the GaAs/Fe$_5$/GaAs system, the interface shows a polarization of 33.8\,\% while the polarization for the second layer is 59.0\,\% and therefore close to the bulk value of 57.7\,\%,\cite{intermag} see center of Fig.\,\ref{fig:dos}.\\
Besides this important variation at the Fermi level, further differences in the DOS exist between Fe atoms at the GaAs(110) surface and bulk Fe. The e$_g$-peak around -1\,eV is shifted to lower energies at the interface, as well as in the second Fe layer in case of a GaAs/Fe$_3$/GaAs slab. In contrast to that, the e$_g$-peak is already at its bulk position in the second Fe layer in case of GaAs/Fe$_5$/GaAs. This difference between slaps of different thickness is related to the relaxation of the ions. Due to a small relaxation of the Fe atoms in direction of the GaAs interface, the atomic volume in the center layer of the GaAs/Fe$_3$/GaAs system is artificially increased. Thus, the difference between the atomic volume of bulk Fe and the increased volume in the supercell leads to a shift of the e$_g$-peak, while this finite-size effect is reduced in case of GaAs/Fe$_5$/GaAs, see also Ref.\,\onlinecite{icm}. The same effect leads to the difference in the layer-resolved polarization between GaAs/Fe$_3$/GaAs and GaAs/Fe$_5$/GaAs. Note that the DOS and thereby $P$ is very sensitive to small changes in the atomic configuration whereas the energy surfaces in the previous sections have shown no variation when using an increased supercell. Also, no significant modifications of the Fe magnetic moments at the GaAs interface appear with increasing thickness of the Fe layer.\\ 
In addition, the GaAs(110) interface leads to a reduction of the split \emph{d}-peak with t$_{2g}$-symmetry around -3\,eV which, however, has no influence on the transport properties in case of moderate voltage.
In summary, one may say that the GaAs interface leads to a small reduction of the Fe polarization at the interface but keeps a quit large polarization. So this system is of interest for spintronic devices.

In case of 1/4 monolayer of Fe on the free GaAs surface,} {\color{black} the modification of the DOS is drastically different. Here, we focus our discussion on configuration "a" in Fig.\,\ref{fig:blubb} for which an Fe atom is at a Ga position in the semiconductor, while a Ga adatom is formed.
In this case, the Fe atoms possess a GaAs-like chemical environment without further Fe atoms in the neighborhood. This type of environment leads to strong modifications of the DOS, see bottom of Fig.\,\ref{fig:dos}. 
The most important modification of the DOS is a large peak in the minority spin channel at the Fermi level.
As the t$_{2g}$-majority-peak at the Fermi level nearly vanishes at the same time, a polarization of -53.4\,\% is obtained at the interface, i.e., the sign of the polarization is reversed  in comparison to bulk Fe. The same reversed sign of polarization (due to an enhanced minority DOS) has been obtained for Fe adatoms on a free GaAs(110) surface \cite{intermag}; thus this seems to be a common feature of highly intermixed  Fe/GaAs interfaces.
In analogy, a large minority peak at the Fermi level caused by localized Fe-\emph{d} states has been found for different interface configurations in (001)-direction.\cite{demchenko:115332} 
The reversal of polarization at the interface may oppose spintronics applications as the DOS will approach its bulk value after some monolayers  if additional layers of Fe are grown on the highly intermixed interface. Thus, mainly majority electrons may be injected from the bulk like Fe layers far from the interface but no corresponding states are available at the interface.  
Also, surface induced minority states may tunnel through the barrier and thereby reduce the polarization of the current through the Fe/GaAs/Fe system.\cite{demchenko:115332}
Among these important modifications of the DOS at the Fermi level, the t$_{2g}$-DOS between -4 and {-2}\,eV approximately increases by a factor of two for the highly intermixed interface.  Additionally, the minority DOS below -1\,eV is strongly reduced. In contrast to this reduction, a peak in the minority DOS appears at  0.5\,eV below the Fermi level.
The whole density of states is reduced, as Fe charge is transfered to the Ga adatoms. 
A Bader analysis \cite{bader_imp} provides a charge transfer of approximately 0.2\,electrons per Fe atom, which is in good agreement with results for the Ga terminated Fe/GaAs(001) interface. \cite{demchenko:115332}

Besides the modification of the Fe DOS, also the GaAs DOS is modified at the interface. In analogy to the Fe/GaAs(001) interface,\cite{demchenko:115332} we found interface induced states at the Fermi level which are highly polarized and decay with increasing distance toward the metal interface, for details see  Ref.\,\onlinecite{icm}.}}
\section{Conclusions and Outlook}
We have investigated the interface structure of different Fe/GaAs(110) configurations with respect to different experimental growth conditions, especially regarding the Fe flux during the deposition. In order to model a moderate Fe flux, we relaxed the free GaAs(110) surface with a quarter of a monolayer Fe adatoms. In a second approach, we investigated the growth process with a larger Fe flux by successively placing Fe layers onto the free GaAs(110) surface. For both cases, we have investigated the structural and magnetic properties of the system.\\
Single Fe atoms on the free GaAs(110) surface lead to a strong relaxation of the topmost GaAs layer and the Fe atoms.
Here, a variety of different interface structures appears, which have similar energy. It is obvious that different experimental studies with similar growth conditions show a wide range of results. Interdiffusion of Ga-Fe atoms through the interface lowers the energy of the interface for low coverages through the under-coordination of the Fe atoms. 
Though, all calculations are performed at $T=0$\,K it is obvious that the relaxation and interdiffusion effects will be enlarged at finite temperatures since the energy differences between the different configurations are very small.\\
Our simulations for a larger amount of Fe on the GaAs(110) surface have shown the formation of flat interfaces without interdiffusion. 
This means that a large Fe flux during the deposition of the first monolayers leads to flat interfaces because the Fe-Fe interaction considerable reduces the interdiffusion. 
This is a hint of how the growth process can be optimized. Since diffusion seems not to be preferable after the growth of 2 monolayers of Fe, the system might be grown at low temperature with large Fe fluxes in order to suppress interdiffusion until a closed Fe layer has formed. These observations are able to explain recent experimental results.\cite{winking:193102,godde}\\
For all investigated interface structures no significant reduction of the magnetic moments appears, the ferromagnetic phase of the Fe atoms was found to be the ground state apart for the case of one monolayer of Fe between two GaAs layers. This latter configuration is however artificial and may only appear locally in case of large intermixing whereas it is very unlikely in case of rather flat interfaces. Therefore, we found the ferromagnetic phase for the (110)-interface to be more stable than in the (001)-interface. In the latter case the formation of highly intermixed magnetic inactive layers often occurs.
{\color{black} Additionally, we found a finite polarization of the density of states even at the direct interface for rather flat interfaces. This behavior of the polarization is advantageous for applications. Opposite to that, a large polarization with an opposite sign appears in case of a highly intermixed Fe/GaAs(110) interface or for (001)-interfaces. The appearance of such large interface states with a reversed polarization may strongly reduce the spin polarized transport.}
 In conclusion, from our calculation it turns out that the Fe/GaAs(110) hybrid system is not only interesting for fundamental reasons but bears potential technological application possibility in future spintronics devices.

\section{Acknowledgments}
This work was supported by the Deutsche Forschungsgemeinschaft (SFB 491 and SFB 445).
\bibliography{anna}

\end{document}